# Why we need all the organisms: an exploration of the Monarch knowledge graph to aid mechanism discovery

Katherina Cortes[1], Daniel Korn[2], Sarah Gehrke[2], Kevin Schaper[2], Corey Cox[2], Patrick Golden[2], Aaron Odell[2], Bryan Laraway[2], Madan Krishnamurthy[2], Justin Reese[3], Harry Caufield[3], Sierra Moxon[3], Ellen Elias[4], Nicolas Matentzoglu, Christopher J Mungall[3], Melissa Haendel[2]

[1] Computational Biosciences Program, University of Colorado Anschutz Medical Campus
[2] Department of Genetics, University of North Carolina at Chapel Hill
[3] Environmental Genomics and Systems Biology, Lawrence Berkeley National Laboratory
[4] Department of Pediatrics and Genetics, Children's Hospital Colorado

## Abstract

Research done using model organisms has been fundamental to the biological understanding of human genes, diseases and phenotypes. Model organisms provide tractable systems for experiments to enhance understanding of biological mechanisms conserved across the evolutionary tree. Decades of model organism research has generated vast amounts of data; however, this data is split across many domains, organisms, and biological fields of research. Knowledge graphs (KGs) are a computational way to aggregate and compile disparate information in a parsable format. By unifying data across studies, organisms and time points, KG researchers can create novel targeted hypotheses. Here we demonstrate how model organisms are connected to humans and other organisms through genes, diseases and phenotypes allowing for a broader understanding of genetic biology than just one organism alone can provide. Utilizing resources such as the Monarch KG is a great way to reduce redundant experiments and find directions previously unexplored.

## Introduction

### Model Organism Use in Disease

Human disorders often have complex origins and varying phenotypic presentation, making it challenging to reveal underlying mechanisms and develop treatments [1,2]. Minor variations in an individual's genome can result in the malfunction of critical biological processes resulting in a genetic/heritable disease [3]. Diseases can also result from modifiable risk factors — environmental toxins, acute trauma, or harmful behaviors such as smoking and drinking — which can dramatically modify an individual's lifetime disease profile and cause non-communicable diseases (NCDs) [4]. To model diseases and understand the mechanistic underpinnings of disease and health, researchers study **model organisms**—non-human species that share conserved biological processes with humans. Compared to stem cell or tissue-based models, whole organisms allow researchers to capture systems-level biology, offering greater experimental flexibility and fewer ethical barriers.

More popular model organisms include the mouse (*Mus musculus*), the roundworm (*Caenorhabditis elegans*), the zebrafish (*Dania rerio*), the clawed frog (*Xenopus laevis*), the fruit fly (*Drosophila melanogaster*), and yeast (*Saccharomyces cerevisiae*). Organisms are chosen to be a "model" based on characteristics like short generation time, ease of care or breeding in the lab, sexual differentiation, historical precedent, and genetic consistency among individuals [5]. Understanding which foundational mechanisms are conserved across species requires piecing together evolutionary clues and leveraging genetic orthology and anatomical homology at different spatiotemporal scales. Throughout the history of investigative biology and medicine, scientists have utilized various animals and plants to apply the findings across the phylogeny of life[6]. Indeed, many notable biological advances would not have happened without animal models: the eradication of polio due to successful vaccine development using mice and primates [7][8], early studies in dogs that demonstrated the role of glucose homeostasis in the role of diabetes [9], enhanced understanding of cell division and human reproductive biology based on research in sea urchins and salamanders [10][11], and the discovery of the mechanism of sex determination using mealworms [12].

Of particular importance is the link between genes and phenotypes which has been the focus of decades of model organism work [13] [14]. Despite these substantive model organism-based findings, only a small fraction of existing genes have been successfully linked to their phenotype [15] [16]. What we have or have not discovered, is a result of researchers being really good at specific types of investigations. The phenomenon of hypotheses being clustered under particular domains is commonly referred to as "looking under the lamp post" [17], which asserts reigning hypotheses and practices, even when incorrect and limiting, dominate investigations and lead to tunnel vision of possible explanations. For example, in 2009, it was reported that "β amyloid, a protein accumulated in the brain in Alzheimer's disease, is produced by and injures skeletal muscle of patients with inclusion body myositis"; however, a bibliographic analysis[18] demonstrated that unfounded claims arose from citation bias.

Arguably, the scientific and medical community has now found everything there is to find in the "genomic lamp light." About a third of the genome is referred to as the "dark genome" due to the lack of understanding of these genes' mechanisms or phenotypes[19,20]. It is probable that many of these genes are responsible for diseases and could be useful druggable targets. Undiscovered gene functions remain, not because we haven't gotten to them yet, but because they remain intrinsically cryptic to prevailing methodologies.

In a 'classic' animal model, the knockout of a human disease gene ortholog, a pair of genes with the same function across different species, will clearly recapitulate many features of the disease; for example, a mouse *CFTR* knockout will show thickened respiratory system mucous secretions leading to respiratory issues, as in cystic fibrosis[21]. However, biology works through the interplay of many genes in the context of the environment and complex causal chains, so model-disease relationships are often less straightforward. This does not imply that model systems are less relevant – in contrast, it opens the opportunity to dissect the mechanisms of pathogenesis more closely. For example, a homozygous knockout of *Ednra* in mouse (MGI:1857473) is known to recapitulate velocardiofacial syndrome (OMIM:192430) associated with variants in *Tbx1[22]*. While these genes are not orthologous, the mechanism can be revealed by zebrafish studies that show that *Ednra* is regulated by *Tbx1* in the relevant pharyngeal precursor [23]. The ability to capture precise phenotypic presentation in both the model organism-human context is vital to understand the nature of the disease recapitulation, as is being able to reason over these relationships using multiple lines of evidence. However, it is difficult for any single researcher to recall all of the necessary knowledge, which becomes even more challenging in the face of modern technologies such as multi-omics characterizing responses at the single-cell and molecular level.

**Inferring new knowledge from existing data**

Swanson described the issue of "Undiscovered Public Knowledge" in various publications [24,25]. In his work, Swanson supposes that knowledge can be broadly divided into two camps; "known knowns", which are facts that have been proven through observation or logical methods and "known unknowns", which are statements that have yet to be proven or disproven (e.g. hypotheses). Swanson argues that by leveraging the state of current information, we become able to derive novel facts through the combination of a variety of seemingly unrelated facts. By using this philosophy, Swanson uncovered a previously undiscovered linkage between magnesium, calcium channels, and migraines [26]. Magnesium remains one of the front-line therapies for migraines today[27]. Translational medicine takes this idea of 'Known unknowns' and seeks to bridge biological fields and their historically siloed information. [28]. Scientists are facing a deluge of data and it is not enough to aggregate the data from heterogeneous sources, one must also annotate, connect, and analyze these data [29]. This level of data generation and knowledge increases the "known unknowns" any given researcher or clinician must understand. Due to the fragmented and siloed nature of model organism data, it is difficult to integrate findings across organisms and experiments [30].

Knowledge graphs (KGs) are a means to navigate and integrate numerous datasets and knowledgebases across the inherent interconnectivity of a domain. Due to fundamental incompatibilities of various perspectives, an essential aspect of this approach is data harmonization-where concepts from different domains are merged into a single conceptual entity. [31] Biomedical KGs often include ontologies, which provide a semantic hierarchical structure for integrating data, such as genes, traits, and phenotypes,[32–34]. Several ontologies are critically relevant to the task of integrating knowledge across species: Uberon, a cross species metazoan anatomy ontology that classifies organs, tissues, and cells via the integrated Cell Ontology (CL); and the Unified Phenotype Ontology (uPheno), which aligns 12 species-specific phenotype ontologies using shared term templates (i.e. Dead Simple Design Patterns) using Uberon, CL, the Gene Ontology, etc. [35]. Through this approach, human phenotypes from the Human Phenotype Ontology (HPO) [36]can be systematically linked to model organism phenotypes, such as those in the Mouse Phenotype Ontology (MP).

## What makes a model organism a model of disease?

There are several ways that a model organism can be considered a model of disease, four of which we include in the Monarch KG. First, a researcher can assert that a given model organism (genotype) is a "model of" a disease [37]. Second, there are orthology relationships where the same gene has a pathological variation in human and the model species. Third, one can compare a disease with a model organism based on phenotypic similarity [38–40]. Finally, one can compare cross species phenotypes by using enrichment of orthologous genes shared between the two phenotypes - termed phenologs (**Fig 1**)[41,42]. While modeling disease can also be performed at the pathway or mechanism level, the four aforementioned methods are most commonly used. Pathway, chemical reaction, and protein-protein interaction data can then be layered upon these four main types of relationships in a KG. Humans have been using other organisms to model disease for centuries; and researchers often assert a "model of" disease relationship based on some aspect of disease or mechanistic resemblance. For example, a researcher might randomly make mutations in the animal's genome to create many animal models, these are then screened to pick ones with similar phenotypes to human disease[43]. In some cases an animal is known to exhibit disease phenotypes similar to that of a human's for example, epilepsy is naturally occurring in dogs and thus is often used as a model for epilepsy even though the molecular mechanism may not be known[44].

One of the most common ways to realize a model association is through orthology, where a known disease gene is mutated in a model species and the outcome compared to the human disease. Orthologs often retain the same function across species due to evolutionary descent from a common ancestor [45], and paralogs are genes within the same organism that have duplicated and often retain related or complementary functions[46]. Model selection can be informed by orthology, and experimentally validated using human genes to rescue a knockout phenotype in organisms like yeast or zebrafish (e.g., BRAF). The best methods for identifying orthologous genes have been a subject of much debate[47], with a spectrum of tools available such as the Protein ANalysis THrough Evolutionary Relationships (PANTHER) classification system [48] and the Swiss Orthology resources[49].

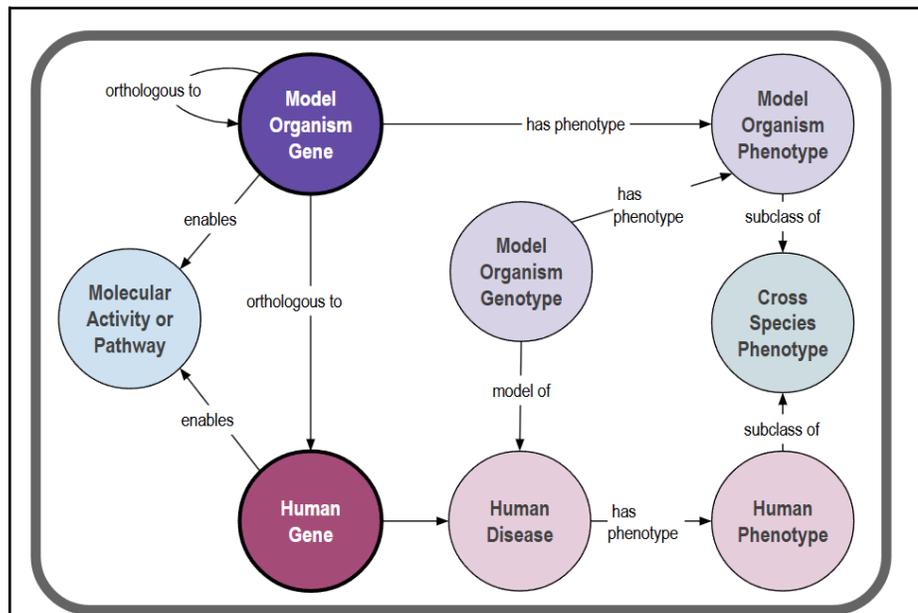

**Figure 1 Overview of Organism Nodes and Relationships in Monarch KG.**
*Graphical overview of how model organism genes, phenotypes, genotypes are connected to other node types in the Monarch KG. This shows a brief overview of how a model organism being a model of disease can be represented. "Model of" and "orthologous to" are both edges in the graphs that are a direct 1:1 relationship between either a genotype and disease or human gene to model organism gene.*

Animal models of disease are created and studied in order to gain a better understanding of the underlying mechanisms and genetics of the disease in humans to ultimately develop diagnostics, therapeutics and targeted drugs[50]. Below, we explore the knowledge available within the Monarch Initiative KG[51], offering a window into "known unknowns" regarding the relationship between model organisms, genes, and phenotypes. The goal is to aid researchers in identifying or creating the right model for the right disease application, based upon the documented, known, and integrated phenotypic diversity across species.

# Methods

### Source data and ontologies used to build the Monarch KG

The Monarch KG has been constructed to bridge the gap between genetic variations, environmental determinants, and phenotypic outcomes in biomedical / biological data in order to support clinical diagnosis and disease mechanism understanding.

The Monarch KG is publicly available in various formats, including Neo4j, TSV, KGX and DuckDB[51]. We used a combination of Neo4j-accessed via Cypher queries and DuckDB-accessed via SQL queries in order to analyze different patterns in the graph. These queries can be accessed in the queries file on the github repo. Our numbers, analysis, and figures are based upon version 0.5.0 from February 17, 2025 (https://data.monarchinitiative.org/monarch-kg/2025-02-17/index.html).

The Monarch KG is composed of two main components; the data and the ontologies that provide a semantic hierarchical schema for the data. The Monarch KG is constantly evolving to fit the needs of the changing landscape of biomedical science and human disease research. In this version, there are over 30 different data sources including: data on model organisms (shown in *Table 1*), human diseases (i.e. MONDO, CTD), protein association networks (i.e. STRINGDB), and phenotype data (i.e. HPO annotations, Alliance Phenotype). Integrating these data sources results in a total of 1.12 million nodes (divided into 17 node types), and 8.91 million edges (divided into 32 edge types), source specifics can be found on the Monarch Initiative website. The Monarch KG was constructed to use and follow the Biolink model specification, which specifies exact domains and ranges of nodes for specific edge types such as *biolink:Causes* linking chemicals/genes to diseases [52]. The Biolink model is used to create biomedical knowledge graphs in a unified way bridging data from disparate sources, such as ontologies, schemas, biomedical databases, and data models.

For the purposes of this evaluation, we evaluated the relationships between model and human genes, phenotypes, biological functions (Gene Ontology), and human diseases within the graph.

The analysis was split into two categories: (1) Analyzing the graph for how cross-species interact by phenotypes only and (2) by all relationship types including phenotypes, mechanisms 5s merging or normalization of nodes by careful choices of standard node ingest sources. Human genes are only from HGNC. Model organism genes are always using the MOD database identifiers. Association ingests that use identifiers outside of the reference node sources are "rewired", in that the sources are read from SSSOM to gather equivalent identifiers and preferred IDs are swapped in (retaining the IDs in alternate fields). Normalization QC is accomplished through a dashboard (http://qc.monarchinitiative.org) that tracks the success rate of normalization for each source ingested.

QC for compliance to the Biolink Model is handled when files are converted to the model in our ingest process. The Biolink Model produces a Pydantic representation of each entity or association in the data model using LinkML Pydantic generator tooling. Each entity or association is validated when it is harmonized to ensure that all fields used are present in the model. Quantitative QC is performed using a yaml configuration that defines expected numbers of nodes or edges in the graph from a given source [34,53].

## Characterizing phenotypic breadth and depth of models and diseases

We first investigated how well the phenotype ontologies were connected in the graph. First, we assessed the number of phenotypic features in each relevant ontology. The number of phenotype nodes from each ontology was retrieved using a neo4j cypher query to extract all biolink: Phenotypic feature nodes in the graph and count based on their type (results in Figure 2).The overlap of human protein coding genes with phenotypic understanding of both the human and orthologous genes was calculated using DuckDB queries.

We next assessed how phenotypes across species were related within the Monarch KG. Neo4j queries were used to count the number of paths between organisms' phenotypes, including human, with a uPheno node in the middle. Graph pattern matching was used to determine how these phenotypes were related through uPheno nodes and edges; *model organism phenotype-> uPheno -> mode organism phenotype (or human phenotype)*. These phenotype paths also looked at connecting phenotypes from the same species such as *mouse phenotype -> uPheno -> mouse phenotype*. All model organism nodes connected to a uPheno node by a *"biolink:subclass_of edge"*. The number of phenotypes that were a part of different phenotype system categories under high level uPheno were investigated. The terms used as the top level terms for each phenotype category can be found in *supplemental* materials.

## Characterizing disease - gene - phenotype relationships through orthology

As most human protein coding genes have orthologs, there was interest in further analyzing the number of orthologs from each species in the graph as well evaluating other types of gene relationships. Namely, we aimed to determine if orthology indicated a likelihood of the human gene having associated diseases, phenotypes, and/or mechanistic annotations.

Our analysis looked at all human genes with orthologs regardless of protein coding designation. We categorized genes on whether they had more than one ortholog present in the graph and whether they were from the same or different organisms. As we are primarily interested in how orthologs might affect disease and phenotype relevancy, we first counted number genes in each ortholog category (one ortholog edge, many ortholog edges from the same organism, many ortholog edges from different organisms). We then calculated the number of genes in each ortholog category that had disease annotations, and finally each that had phenotype annotations. This analysis was intended to illuminate how orthology might inform disease and phenotypic understanding.

The graph was also queried to see if certain pairs of organisms were more likely to have orthologs than other pairs. For example, do rat and mouse have more orthology edges represented in the KG than rat and clawed frog since they are more closely related on the evolutionary tree of life? Conversely, do some model organisms have a larger representation of orthology due to biological research bias? To inspect orthology edge quantity across different pairs of organisms, we pulled all "*orthologous to*" edges from the graph and recorded the number of times an ortholog appeared between a set of organisms. As there are currently no paralogous edges in the graph, there will not be orthologous edges between two nodes of the same organism, they will however be assigned to the same protein family in Panther. The number of orthology edges for each model organism was assessed followed by the total number of human protein-coding genes, identifying how many had orthologs and how many did not.

## Querying the graph to reveal knowledge that can be gained through cross species interactions

Based on the graph interrogations above, we then used neo4j to find examples for some of the aforementioned categories. The three main types of information that orthologs can provide to human genes is protein-protein interactions, phenotypes and go terms. For each of these types, an example was selected where the ortholog would provide new information about the human gene.

Ehlers Danlos Syndrome (EDS) is a heritable connective tissue disorder with 14 subtypes. Along with affecting skin and wound healing, EDS often presents as various medical problems such as cardiological, neurological and gastrointestinal disorders. Focusing on one of the subtypes of EDS, hypermobile EDS (hEDS), which is thought to affect 1 in 5000 people in the world and 80-90% of EDS cases. Despite its prevalence, there is no known gene or underlying mechanism associated with hEDS. There are thought to be as many as 12 genes involved; however, the lack of knowledge about the underlying mechanism creates uncertainty about the role of the genes that may be potentially involved. The variation of symptoms and genes associated with hEDS make it difficult to diagnose and treat. Using a subset of genes thought to be implicated in hEDS, provided by a clinical geneticist who specializes in EDS, we searched for phenotypically similar genes to widen the search for causal genes. We used Phenodigm and jaccard similarity to investigate genes for phenotypic similarity to the implicated gene list.. Phenotypes of genes presumably implicated, i.e., TNXB, MYLK, COL5A1, MYH11, COL12A1, COL1A1, COL1A2, were expanded using mouse orthologs. hEDS-associated genes and phenotypes were used as input to semantic similarity algorithms (Jaccard and Phenodigm). Semantic similarity measures the similarity between sets of terms based on their meaning. Twenty (20) genes were selected from each method using each input, 80 total.

To illustrate how phenologs help to bridge the gap between human phenotypes and other model organism phenotypes using graph connections, we selected the highest scoring phenolog between human and xenopus phenotypes.

## Results/Case Study

In the Monarch KG, there are many represented organisms. **Table 1** displays the common name of each organism along with the scientific name, NCBI Taxon, data source, number of gene nodes and edges from those nodes present in the graph.

*Table 1 Organism Types, Sources and Counts*
*Model organisms, node counts, number of connections in graph, source of data.*

| Common Name | Taxon | NCBI Taxon ID | Gene Count | Edge Count | Data Source |
|---|---|---|---|---|---|
| Humans | *Homo Sapiens* | NCBITaxon:9606 | 44454 | 4401949 | HGNC |
| Jungle Fowl | *Gallus gallus* | NCBITaxon:9031 | 32180 | 528887 | NCBI |
| Cow | *Bos taurus* | NCBITaxon:9913 | 57279 | 836607 | NCBI |
| Fungi | *Aspergillus nidulans FGSC A4* | NCBITaxon:227321 | 10569 | 204189 | NCBI |
| Dog | *Canis lupus familiaris* | NCBITaxon:9615 | 50756 | 586670 | NCBI |
| Mouse | *Mus musculus* | NCBITaxon:10090 | 97469 | 3004611 | MGI (MGD) |
| Fly | *Drosophila melanogaster* | NCBITaxon:7227 | 30244 | 1171992 | FlyBase |
| Zebrafish | *Danio rerio* | NCBITaxon:7955 | 37941 | 1761988 | ZFIN |
| Frog | *Xenopus tropicalis* | NCBITaxon:8364 | 15746 | 374957 | Xenbase |
| Rat | *Rattus norvegicus* | NCBITaxon:10116 | 61335 | 1247786 | RGD |
| Worm | *Caenorhabditis elegans* | NCBITaxon:6239 | 48777 | 783831 | WormBase |
| Soil dwelling amoeba | *Dictyostelium discoideum* | NCBITaxon:44689 | 14222 | 220722 | dictyBase |
| Fission Yeast | *Schizosaccharomyces pombe* | NCBITaxon:4896 | 5134 | 381359 | PomBase |
| Frog | *Xenopus laevis* | NCBITaxon:8355 | 22992 | 70205 | Xenbase |
| Budding Yeast | *Saccharomyces cerevisiae S288C* | NCBITaxon:559292 | 7167 | 247019 | SGD |
| Wild Boar | *Sus scrofa* | NCBITaxon:9823 | 45437 | 508961 | NCBI |

### Table 2. Human Protein Coding Genes and Annotation Types

*Number of human protein coding genes with different types of annotations. Annotation types are: protein-protein interaction, phenotype or biological activity. Annotation is only counted if it is represented as an edge in the Monarch KG. For each annotation type there are three additional parameters: has annotation, has ortholog and has ortholog annotation. These additional parameters indicate if the gene has the annotation type, an ortholog and if the ortholog has the annotation type. Each gene is represented in each annotation type, they are not exclusive.*

| Annotation Type | Count | has_annotation | has_ortholog | has_ortholog_annotation |
|---|---|---|---|---|
| protein protein interaction | 17716 | ● | ● | ● |
| | 511 | ● | ● | |
| | 445 | ● | | |
| | 409 | | ● | ● |
| | 170 | | ● | |
| | 200 | | | |
| phenotype | 5174 | ● | ● | ● |
| | 209 | ● | ● | |
| | 22 | ● | | |
| | 10504 | | ● | ● |
| | 2919 | | ● | |
| | 623 | | | |
| biological activity | 18334 | ● | ● | ● |
| | 90 | ● | ● | |
| | 332 | ● | | |
| | 318 | | ● | ● |
| | 64 | | ● | |
| | 313 | | | |

Orthologs are used to garner information about less well characterized human genes. As shown in **Table 2,** there are human protein coding genes with no associated protein protein interaction, phenotype or biological activity where the ortholog has that type of annotation. There are 401 human protein coding genes with no protein protein interaction annotation but the ortholog has a protein protein annotation. Similarly, there are 10,529 genes with no phenotype annotations that have an ortholog with phenotype annotations. Finally, there are 331 genes with no biological activity annotations that the ortholog has biological activity annotations. Most genes have biological activity or protein protein interaction annotations but fewer have phenotypic annotations.

A simple neo4j query revealed that there are 2,066 human diseases in the graph with "*model of*" relationships to a model organism genotype. Of these 2,066 diseases 1,382 have human gene connections, meaning there are 684 human diseases for which biologists have a model organism used to study the disease without having a known disease causing human gene.

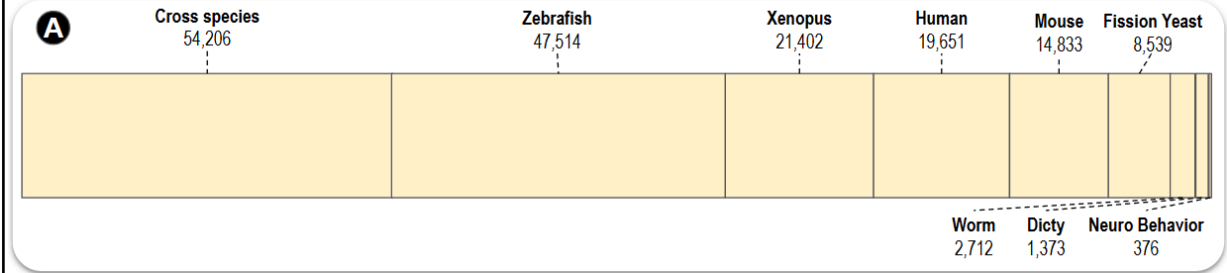
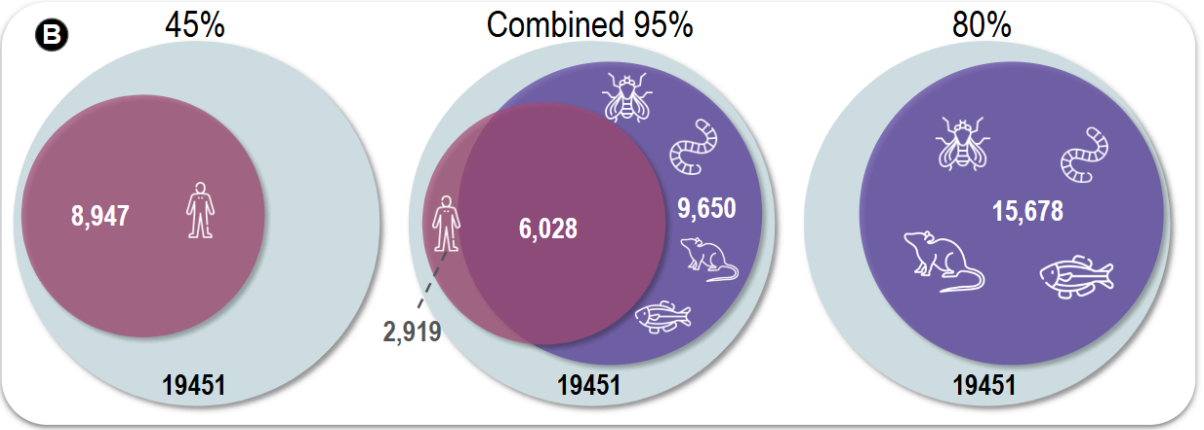
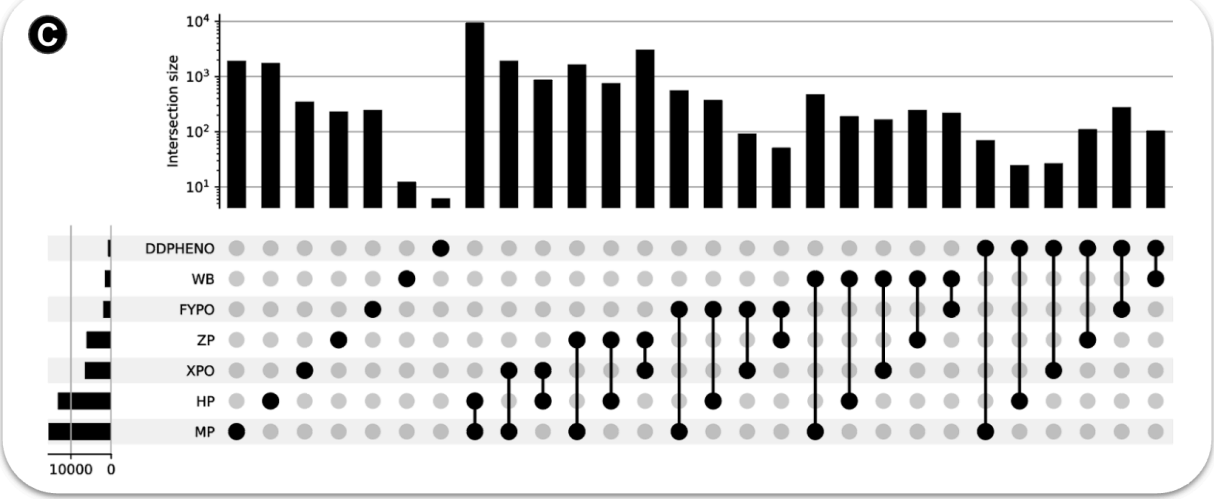

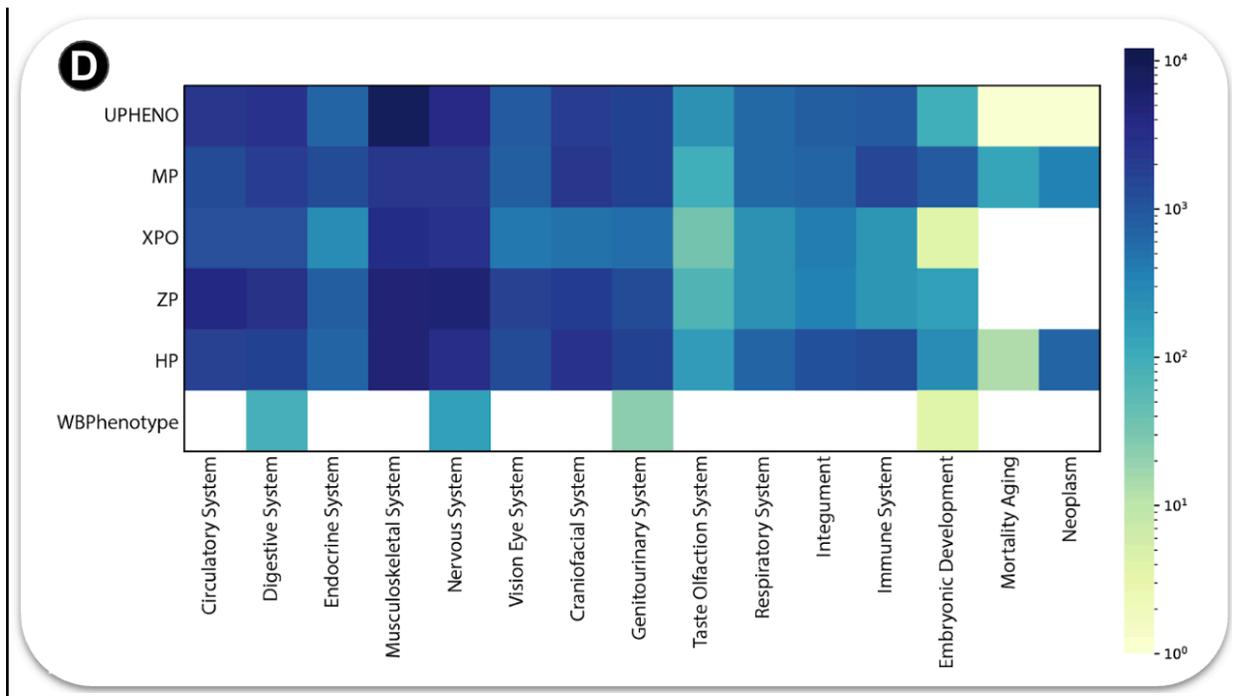

*Figure 2A-D. Phenotypic characterization of models and diseases in the Monarch KG.*
*A. Number of phenotypic features in each phenotype ontology. Size of rectangle corresponds to relative amount of phenotypic features as compared to the total number in all ontologies. B. Venn Diagram of human protein coding genes in light blue (19,451) with number of genes with associated phenotypes in pink (9,739),compared to number of human genes with orthologs that have associated phenotypes (14,779). The middle circle is the combination of human genes with phenotypes and orthologs with phenotypes. C. Upset plot that shows how different species specific phenotype ontologies are connected by upheno nodes. D. Heatmap of phenotype categories that are a subtype of more general upheno categories. These four panels show how ortholog phenotypic information contributes to human gene understanding.*

Further investigation into these 684 diseases with *"model of"* relationships but not human gene relationship reveals that most of these diseases, 560, are larger classes that have other diseases as subclasses. These super classes of disease would be unlikely to have human gene associations anyways. In addition, 10 of the 684 disease terms are obsolete, meaning no longer used or supported by the ontologies and databases, thus not expected to have many annotations. Removing the disease superclasses and obsolete terms there are 114 diseases without human gene associations that we might reasonably expect to have associated genes.

There are nine different ontologies that contribute phenotypic feature nodes to the Monarch KG, the number of each is shown in **Fig 2A**. Cross Species phenotypes (UPheno ontology) has the most phenotypes in our graph followed closely by Zebrafish. The total number of human phenotypes, in comparison, have about half of the two most abundant phenotypes in the KG. The Neuro Behavioural Ontology (NBO) has the least phenotypes followed by the less biologically complex organisms, slime mold, worms, and yeast. While commonly used model organisms in the research community, budding yeast, fly, and rat are not represented by phenotypic ontologies as shown in **Fig 2A**. Similarly, there are no phenotypes associated with the model organisms coming from NCBI, jungle fowl, dog, cow and pig. The phenotypes in **Fig**

**2A** overlap and provide rich information to human genes as shown in **Fig 2B**. Cross species phenotypes enriches our understanding of human genes by utilizing all the information from these other species. *D*ifferent phenotype ontologies are connected through UPheno nodes (**Fig 2C**). The organisms that share the most phenotypes are also the most phylogenetically similar. For example, zebrafish and xenopus have the highest shared phenotypes followed by mouse and human. Fission yeast on the other hand shares relatively few phenotypes with other organisms (**Fig 2C**). Different model organisms also have different kinds of phenotypes as shown in **(Fig 2D)**: worms only have musculature, digestive and nervous system phenotypes, while more complex organisms like zebrafish and mouse have all types. The amount of phenotypic terms relate to the top level terms of the uPheno ontology; so there are some phenotypes which map to multiple top level terms. In addition, there are some phenotypes that do not map to any of the high level terms here. For example, in the cases of both Zebrafish and Xenopus, there are neoplasm and embryonic phenotype terms which do not map to the high level uPheno neoplasm or embryonic phenotype term. Most protein coding genes have GO annotations and protein protein interactions documented in the graph. There are a lot less human genes with phenotypic information.

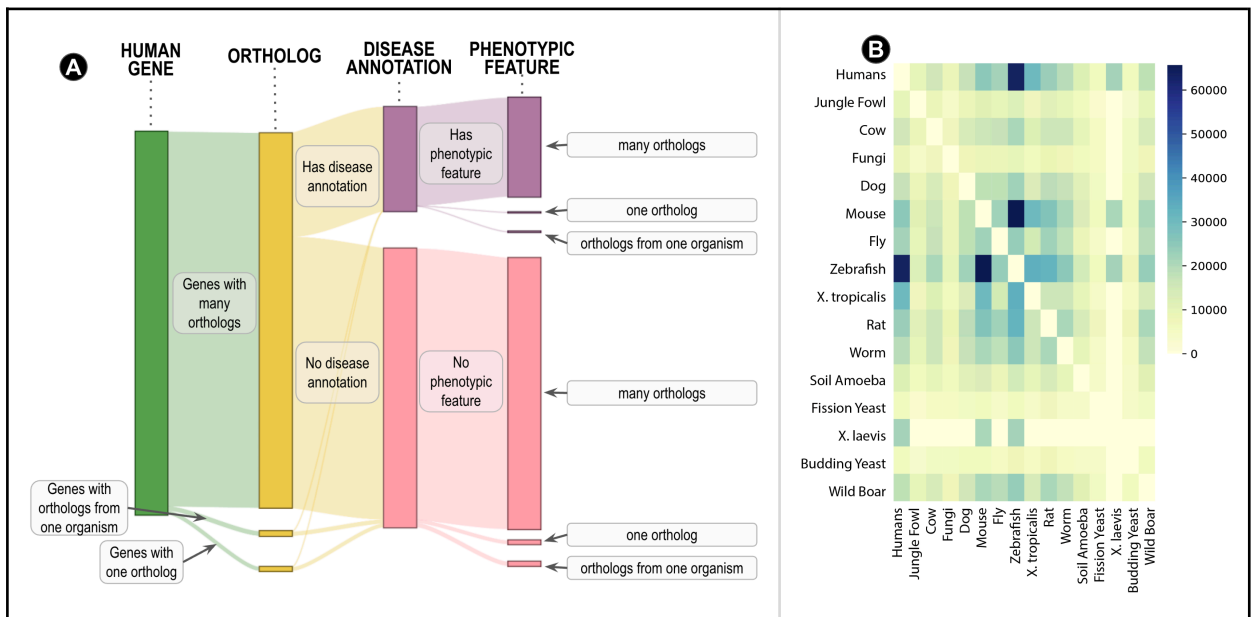

*Figure 3. Orthologs Across Organisms and their influence on Human Gene Annotations.*
*A: Sankey all human genes regardless of if they are designated as protein coding or not B: Heatmap detailing numbers of orthologs between species. The two panels show how use of cross species orthologs contribute to human gene annotations*

The connections between cross-species orthologs is illustrated in **Fig 3**, with additional information on how orthologs are related to phenotypes and diseases in humans **(Fig 3A)**. As shown in **Fig3A,** most human genes with ortholog and disease annotations also have phenotype annotations; similarly, human genes with orthologs but without disease annotations usually do not have phenotype annotations. And, most human genes have many orthologs from different species. Only 198 genes have one ortholog and only 233 human genes have more than one ortholog that all come from the same species. By interrogating the Monarch KG to assess if some organisms have more orthologs from one organism than others, it was found that there are many different orthologs between different species in the graph with a few exceptions **Fig 3B**. As orthologs are genes from different species with the same function there are no orthologs between the same species for example, there are no mouse to mouse orthologs. Those would be considered paralogs which are not represented in our graph. Jungle fowl and fungi had no ortholog connections in the KG. There are two different frog species in our graph xenopus laevis and xenopus tropicalis. Due to X. laevis being a tetraploid species, the genome and transcriptome annotations are not as good as other species, because of this there are less orthology connections. X. tropicalis has orthology connections with all other species that have orthologs except the other frog. The X. laevis species only has ortholog interactions with humans, mice and zebrafish. This is due to the tetraploid nature of the X. laevis genome, these genome and transcriptome annotations are less robust and therefore often excluded. The species with the highest amount of orthologs was zebrafish -human and mouse-zebrafish. While rat and mouse had a high number of orthologs it was not the highest of the pairs.

**Case Study- Inferred from Mutant Phenotype (SNORD118)**

Orthologs can help give important information in areas where there is little. For example, as shown in **Fig 4A** SNORD118 gene is a small nuclear RNA in humans that is known to cause leukoencephalopathy with calcifications and cysts and interacts with genes PHAX and TRIM25. SNORD118 also has some associated phenotypes and anatomical expression locations. With the addition of snord118a, an orthologous gene in zebrafish to SNORD118, it is inferred from the Monarch KG that this gene acts upstream of or within ribosomal small subunit biogenesis. **Fig 4B** highlights the important connections that lead to this inference. Ribosomal small subunit biogenesis is the only cellular process/GO term connected to this gene in our graph. In addition to being able to give a starting point for novel mechanistic experimentation, this information also provides a potential model organism candidate for future biological studies. The edge between snord118a (zebrafish) and ribosomal small subunit biogenesis was asserted in the KG using an *"inferred from mutant phenotype"* evidence. Further investigation shows that this edge is supported by a paper looking into how mutations in this gene cause LCC[54].

SNORD118a is not a protein coding gene but orthologs can help us to further investigate these more unusual types of genes that often have less associated information due to their unusualness. Furthermore, this is a great example of how model organisms could be better utilized to approach the "dark genome", snord118a encodes for an RNA U8 and yet has a demonstrated connection to a disease. However, SNORD118 orthologs have been better aligned to the human gene in mammals than in zebrafish or xenopus; it is possible that we would be able to discover even more about leukoencephalopathy molecular pathology.

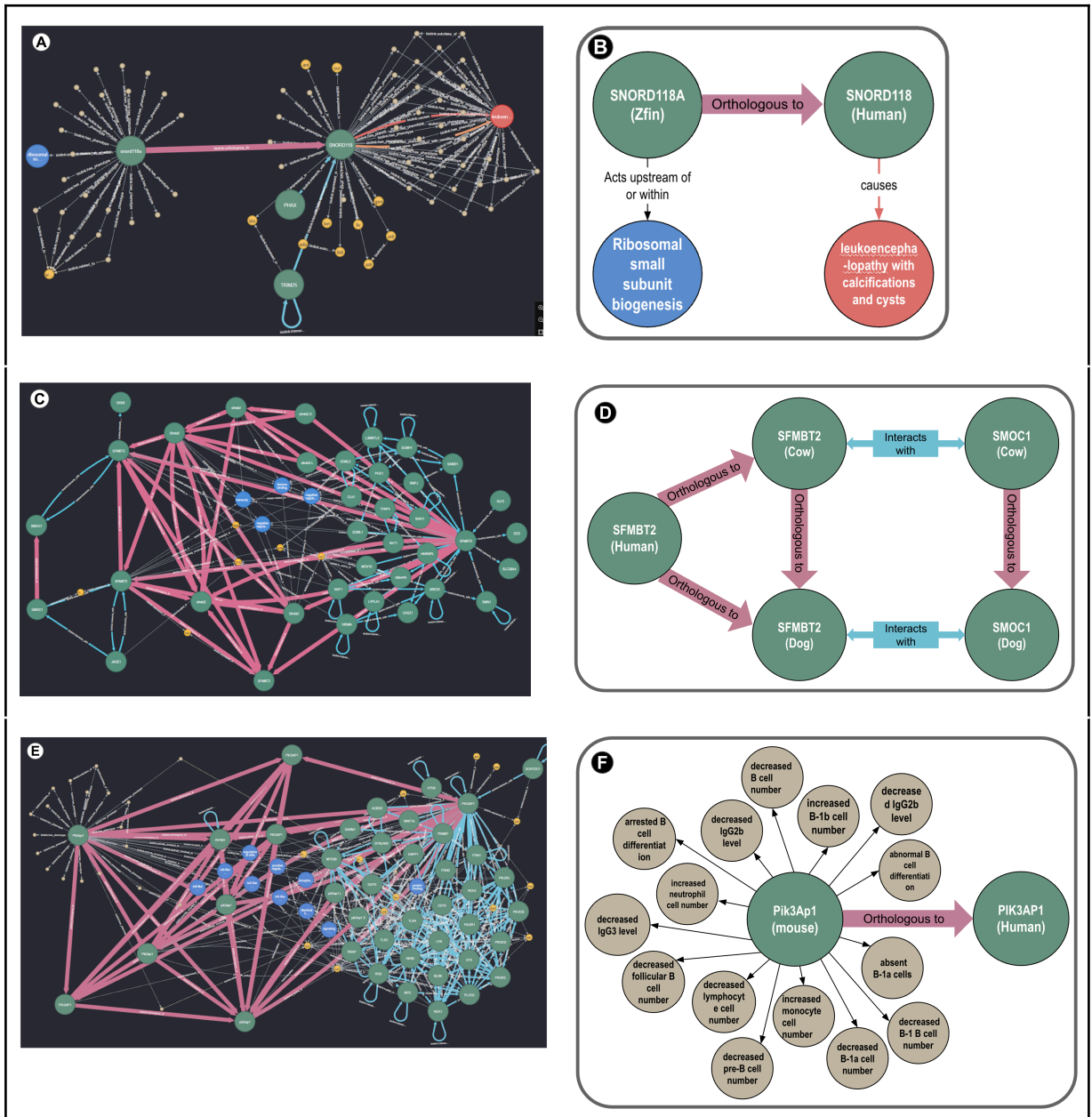

**Figure 4A-F Monarch KG Examples of Orthologs Providing New Information to Human Genes**
**A:** Neo4j example of SNORD118 and its direct connections in the Monarch KG as well as the direct connections to snord118a, an ortholog of SNORD118. **B:** zoomed in version of interesting connections to SNORD118 **C:** Neo4j example of SFMBT2 and its direct connections in the Monarch KG as well as the direct connections to the orthologs of SFMBT2 and their protein protein interactions with SMOC1. **D:** zoomed in version of interesting connections to SFMBT2 **E:** Neo4j example of PIK3AP1 and its direct connections in the Monarch KG as well as the direct connections to Pik3aP1, an ortholog of PIK3AP1. **F:** zoomed in version of interesting connections to PIK3AP1

## Case Study - Protein Protein Interaction (SFMBT2)

Another way orthologs could potentially be useful is in finding novel human protein protein interactions. For example, in the case of SFMBT2, a gene which has many protein-protein interactions and orthologs within the Monarch KG, but no disease or phenotypic information. SFMBT2 plays a role in regulating cell growth and development and as such is implicated in different cancers. SFMBT2 has an ortholog in dog as well as cow, shown in **Fig 4C-D**, both of which have protein protein interactions with the gene SMOC1, which has both phenotypic and disease relationships within the Monarch KG. SMOC1in comparison has both associated phenotypes and a disease, microphthalmia with limb anomalies. SMOC1 and SFMBT2 do not have a protein protein interaction in humans; thus this could be an interesting avenue for pursuing new protein protein interactions in future research.

## Case Study - Phenotypes (PIK3AP1)

Orthologs and their related phenotypes provide a great way to get a broader understanding of the gene mutation effects. For example, PIK3AP1 has many associated protein interactions as well as biological processes, orthologs and anatomical entities, it does not however have any associated phenotypes in our graph shown in **Fig 4E-F**. While this means it is easy for researchers to understand its mechanistic function and place in biological pathways its possible disease presentation remains unclear. The phenotypes associated with the orthologs could give researchers a preview into what the gene might do on a broader scale.

There are multiple ways that having model organism phenotypes could be helpful in disease - gene discovery. Using the Monarch KG's phenotype analysis tool one can get potential diseases that have similar phenotypic profiles similar to the ortholog gene's phenotypes. This tool could be used to generate hypotheses about what disease a mutation in the gene would cause. Conversely, if one was trying to find the causal gene of a specific disease this process could be done in reverse, looking for orthologs with similar phenotypic profiles in places where the human disease has no causal gene.

## Case Study - Phenologs

Using the phenologs framework, one can associate cross species phenotypes to human diseases [41]. In **Figure 5**, we show the top significant phenolog associations found across mouse, zebrafish, and pombe in relation to human disease primary ciliary dyskinesia. Leveraging the methodology from Woods et al. 2013 [42], direct orthologous links between cross species genes are collapsed into ortho groups. In this case, ortho groups refer to the protein family assigned to each gene by PantherDB. We are therefore linking cross species phenotypes via the underlying protein association networks. By looking for genes within these networks associated with a given phenotype, but not directly associated with a human disease, one can infer potential novel gene-to-disease candidates. In this specific example, one could hypothesize that genes found to be associated with both hydrocephaly in mouse, and abnormal cystic pronephros in zebrafish could be novel candidate genes for human primary ciliary dyskinesia. Moreover, this information can be systematically computed for all human diseases within the knowledge graph, thus demonstrating a programmatic and high throughput approach for identifying novel human disease candidate genes.

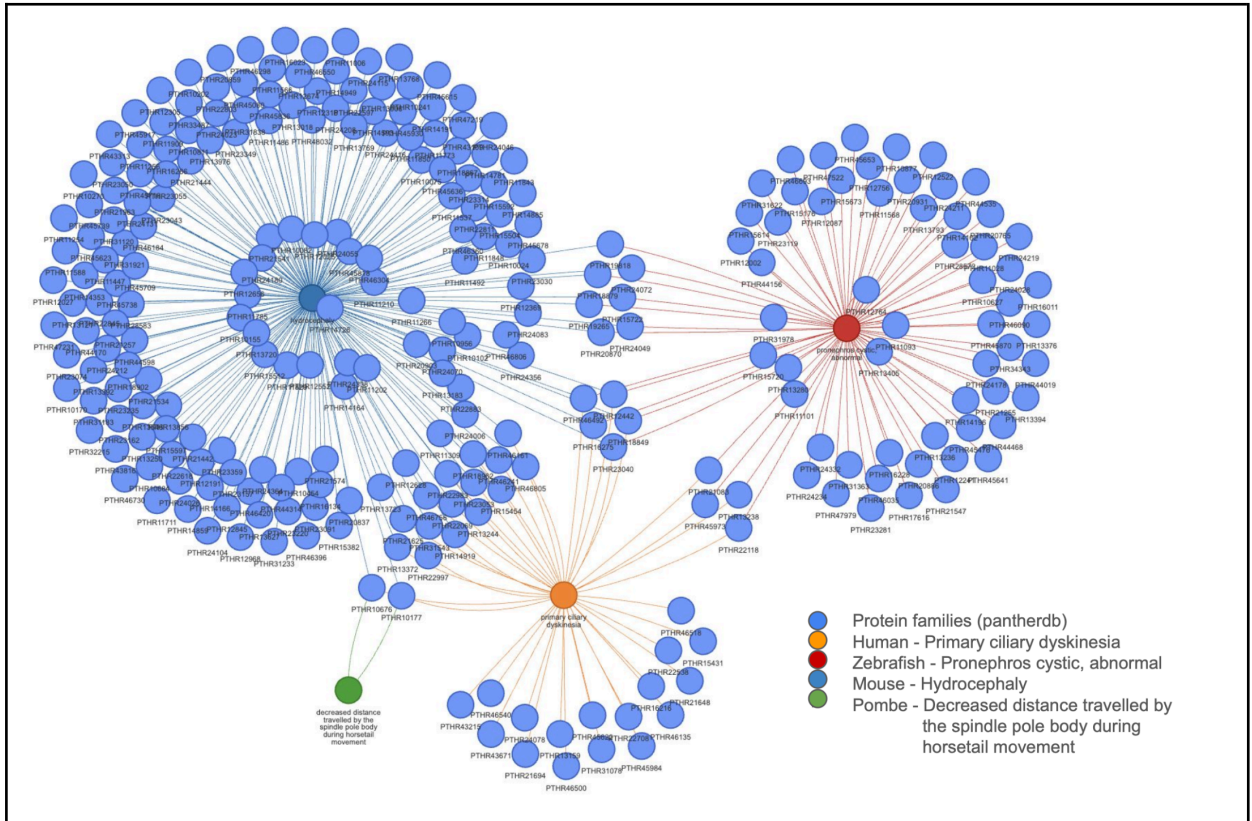

*Figure 5. Phenolog associations across human, zebrafish, mouse and pobme organisms.*
*Blue represents protein families with many being shared across these organisms in relation to the human disease primary ciliary dyskinesia.*

## Case Study Phenotypic Similarity (hEDS)

Many genes were found across all four results, including COL3A1 and PLOD1, which are both known to be associated with other types of EDS. With the goal of finding newly implicated genes, we focused on genes that were not already associated with hEDS or other EDS types, shown in **Fig 6A**. Two of the genes returned were LOX and ATP7A; both are related to copper metabolism. Notably, LOX mutations were found by the clinical geneticist in patients with hEDS. Both genes have phenotypes often seen in hEDS patients including joint hyperflexibility and aortic aneurysm.

Copper pathways have been suggested to play a role in EDS, however these genes had not yet been linked to hEDS. In ongoing work, we are further investigating LOX as a potential key gene in the underlying mechanism of action for hEDS. hEDS Phenotype Similarity

Comparison of LOX and ATP7A phenotypes and their similarities to hEDS phenotypes. The results generated using the Monarch KG are visualized into a grid (Phenogrid). The number above each column and box colors represent how similar the gene phenotypes are to hEDS. Joint hyperflexibility is a symptom commonly seen in hEDS patients, which makes it of great significance to be related to both the LOX and ATP7A genes as shown in **Fig 6B**.

## Discussion/Future Work

Interactions of orthologs provide interesting insights into genes that interact but do not yet have a proven/shown interaction in humans. However, many biomedical model organisms are not included in the Monarch KG due to their lack of data curation. Many genotype-phenotype associations for core model organisms have been curated in Model Organism Databases (MODs), such as Saccharomyces Genome Database (SGD) and Zebrafish Information Network (ZFIN). The Alliance of Genome Resources (AGR) provides common infrastructure for several MODs[55]. These resources provide standardized database entries and ontology terms for genes, genotypes, alleles, phenotypes, etc.[56]. However, a large amount of the literature — including many relevant non-MOD animal models of disease — remains uncurated. Moreover, researchers are often interested in specific phenotypic features (e.g. a craniofacial researcher) and may not look for potentially interesting or additionally relevant phenotypes (e.g. cardiac), leading to bias by failing to consider all of the phenotypic details of a given model. For example, despite being a common model for biomedical studies, rabbit does not have a large curation effort to link it with other model organisms and human diseases. There exists RabGTD which curates information about rabbit genes and transcripts but does not include phenotypic information. This database is not ingested into the Monarch database, yet rabbits have a more similar genome to humans than rodents do and are often used to model cardiovascular diseases, eye diseases, and infectious diseases. The information contained from the long history of rabbit experimentation could provide vital insights to other model organisms, genes or diseases but due to its siloed nature this information cannot be fully utilized.

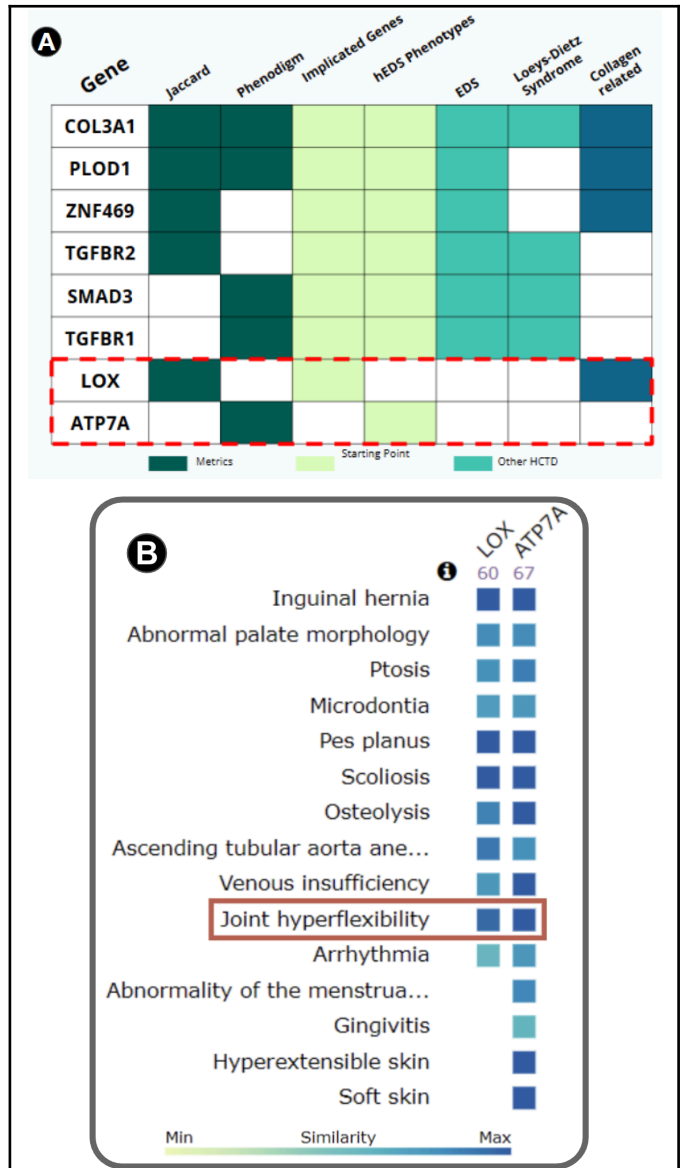

*Figure 6. **A:** Notable Genes. Columns show genes selection method, implicated diseases and if their function is collagen related. Many have been previously implicated in EDS or Loeys-Dietz Syndrome (another heritable connective tissue disorder). LOX and ATP7A (highlighted in red), were not implicated in EDS or Loeys-Dietz Syndrome and both have functions involving copper, which is of particular interest. **B:** Comparison of LOX and ATP7A phenotypes and their similarities to hEDS phenotypes. The results generated using the Monarch KG are visualized into a grid (Phenogrid). The number above each column and box colors represent how similar the gene phenotypes are to hEDS. Joint hyperflexibility is a symptom commonly seen in hEDS patients, which makes it of great significance to be related to both the LOX and ATP7A genes.*

Adding in the ortholog phenotypic information greatly increases the number of human genes with phenotypes; this is very important in the case of disease diagnosis and treatment especially for rare diseases. Just because we know what a gene does mechanistically does not mean that we know what phenotypes will arise on a broader, whole organism level.

Most human genes likely have orthologs in more than one species because when an ortholog is found, it is often studied in multiple species. In addition, genes with disease annotations are likely to have phenotypes as diseases are often characterized by the phenotypes of the patients. If a gene has no disease annotations, it is possible that a mutation of that gene is incompatible with life or is so rare that it has not been seen frequently enough to understand or research it. With the increasing use and accessibility of DNA sequencing it is likely that there will also be an increase in ortholog declaration.

Phenotypic annotations are important for rare disease diagnostics. The human protein coding genes with no annotation and with no annotation on the ortholog could be good places of future study. The orthologs could be interrogated for better understanding of the protein protein interactions or phenotypes or biological activity to bring better understanding to the human gene and its functions and activity.

Due to the evolutionary closeness of rat - mouse and zebrafish - frog, one might expect them to have more orthologs than other organism pairs. However, this illustrates how known knowledge is driven by human choice and bias instead of biology. Unsurprisingly, the yeast models had comparatively few orthologs with other organisms likely due to their lower number of genes. Notably, humans, mouse and zebrafish had the highest number of orthologs, likely due to the popularity of using mouse and zebrafish for models for the purpose of learning more about human biology

Orthologs could help scientists in finding new protein-protein interaction in human genes. Use cases such as the ones shown in **Fig 4** could be used as a jumping off point for researchers to investigate new protein protein interactions. Understanding how the data, such as phenotypes, are not connected, especially in cases where one might expect them to be, is crucial to cross-species research and investigation. Unfortunately, cases where data is expected to be connected can be broken down into two categories; 1) the data does not exist and is something thought to be true but not proven 2) the data does exist but is not connected in the databases being used and as such is shown to not exist.

With growing use of machine learning, LLMs, and high computing clusters there is a growing ability to make use of these vast amounts of information in meaningful ways. The aggregation of model organism data could help researchers and clinicians to elucidate hidden patterns such as rare disease mechanisms, drug repurposing or automated selection of a model organism for a disease. By using the Monarch KG, researchers can narrow down combinations of genes to test in model organisms in the laboratory, thus optimizing both the time and money spent on biomedical experimentation and getting results to real world patients faster. Beyond core model organisms such as the ones shown above in **Fig 2C**, there is no equivalent domain expert-driven phenotype curation present in model organism databases, such as fly, dog, pig or cow. The way phenotypes are recorded across species can also differ depending on database, curation efforts and standards. Due to this lack of standardization and widespread phenotypic curation efforts, a lot of crucial animal genotype-phenotype data may be invisible to computational approaches. For example, transgenic primate models of Parkinson's disease can recapitulate the neurodegenerative patterns of Parkinson's in ways that core model organisms cannot[57]. Another example is the use of CRISPR to create primate models of HIV infection[58]. Non-primate examples include rabbit models of Duchenne muscular dystrophy that better recapitulate features of the human disease[59], and a conditional *CFTR* knockout in ferrets that elucidated molecular aspects of cystic fibrosis[60]. There are many such examples; the number of relevant non-model organism disease models is growing substantially.

Despite the great strides that model organism research allows biologists and researchers to make; a model organism is still just a model. Research must still be done to bring the knowledge from model organisms to human level understanding. There can be a large gap between how a non-human organism responds to drugs or stimuli vs a human one. This analysis is not to suggest that human research, clinical trials and understanding should be replaced by model organism research or that research done on model organisms should be considered any sort of general truth about the universe and state of biology. This analysis is however calling to attention the vast wealth of knowledge we have already obtained using model organisms and how we can better leverage that knowledge to make future targeted hypotheses. The ability to aggregate, disseminate and parse large amounts of data has never been more accessible to the average researcher. As science progresses and most of what is "in the lamp light" has been uncovered; in order to look beyond that light it is necessary to pull from different fields of biology in a way not previously done.

# Supplemental

| Upheno term | Upheno Name | Phenotype category |
|---|---|---|
| UPHENO:0005092 | Musculoskeletal system phenotype | Musculoskeletal system |
| UPHENO:0002816 | Musculature Phenotype | Musculoskeletal system |
| UPHENO:0002964 | Skeletal System Phenotype | Musculoskeletal system |
| UPHENO:0002712 | Connective Tissue Phenotype | Musculoskeletal system |
| UPHENO:0002635 | Integument Phenotype | Integument |
| UPHENO:3000000 | Craniofacial/craniocervical Phenotype | Craniofacial System |
| UPHENO:0002910 | Face Phenotype | Craniofacial System |
| UPHENO:3000001 | Embryonic Development/Birth Phenotype | Embryonic Development |
| UPHENO:0049762 | Embryonic Morphogenesis Phenotype | Embryonic Development |
| UPHENO:3000002 | Mortality/Aging Phenotype | Mortality Aging |
| UPHENO:3000003 | Neoplasm Phenotype | Neoplasm |
| UPHENO:3000004 | Renal/Urinary System Phenotype | Genitourinary System |
| UPHENO:0002832 | Renal System Phenotype | Genitourinary System |
| UPHENO:0002642 | Genitourinary System Phenotype | Genitourinary System |
| UPHENO:0003055 | Reproductive System Phenotype | Genitourinary System |
| UPHENO:3000006 | Taste/Olfaction Phenotype | Taste and Olfaction System |
| UPHENO:0002715 | Olfactory Organ Phenotype | Taste and Olfaction System |
| UPHENO:3000007 | Vision/Eye Phenotype | Vision and Eye System |
| UPHENO:0004757 | Eye Phenotype | Vision and Eye System |
| UPHENO:0004536 | Respiratory System Phenotype | Respiratory System |
| UPHENO:0002833 | Digestive System Phenotype | Digestive System |
| UPHENO:0003116 | Endocrine System Phenotype | Endocrine System |
| UPHENO:9002003 | Circulatory System Phenotype | Circulatory System |
| UPHENO:0002948 | Immune System Phenotype | Immune System |
| UPHENO:0004523 | Nervous System Phenotype | Nervous System |
| UPHENO:0004521 | Central Nervous System Phenotype | Nervous System |

**Supplemental Table 1:** *The Upheno top level terms used for Figure 2D.*